\documentclass[aps,preprintnumbers,amsmath,amssymb,prd,superscriptaddress,longbibliography,twocolumn]{revtex4-1}
\usepackage{dcolumn}
\usepackage{color}
\usepackage{subfigure}  
\usepackage{dblfloatfix}
\usepackage{feynmp}
\usepackage{feynmp-auto}

\def\comment#1{}

\newcommand{\nc}{\newcommand}
\nc{\beq}{\begin{eqnarray}}
\nc{\eeq}{\end{eqnarray}}
\nc{\scs}{\scriptstyle}
\nc{\setval}{\fmfset{wiggly_len}{3mm} \fmfset{arrow_len}{1.5mm}
	\fmfset{arrow_ang}{13} \fmfset{dash_len}{1.5mm}\fmfpen{0.125mm}
	\fmfset{dot_size}{2thick}}

\usepackage{bm,latexsym,mathrsfs,enumerate,color}
\usepackage[mathcal]{euscript}
\usepackage[breaklinks=true,unicode=true,urlcolor = blue,colorlinks = true,citecolor = blue,linkcolor = blue]{hyperref}
\usepackage{graphicx}
\usepackage{todonotes}
\usepackage{wrapfig}

\def\slashchar#1{\setbox0=\hbox{$#1$}           
	\dimen0=\wd0                                 
	\setbox1=\hbox{/} \dimen1=\wd1               
	\ifdim\dimen0>\dimen1                        
	\rlap{\hbox to \dimen0{\hfil/\hfil}}      
	#1                                        
	\else                                        
	\rlap{\hbox to \dimen1{\hfil$#1$\hfil}}   
	/                                         
	\fi}                                         %

\DeclareMathAlphabet\mathbfcal{OMS}{cmsy}{b}{n}

\begin{document}
	
\title{Conformality loss and quantum criticality in topological Higgs electrodynamics in 2+1 dimensions} 
\author{Flavio S. Nogueira}
\affiliation{Institute for Theoretical Solid State Physics, IFW Dresden, Helmholtzstr. 20, 01069 Dresden, Germany}

\author{Jeroen van den Brink}
\affiliation{Institute for Theoretical Solid State Physics, IFW Dresden, Helmholtzstr. 20, 01069 Dresden, Germany}
\affiliation{Institute for Theoretical Physics, TU Dresden, 01069 Dresden, Germany}
\affiliation{Department of Physics, Washington University, St. Louis, MO 63130, USA}

\author{Asle Sudb{\o}}
\affiliation{Center for Quantum Spintronics, Department of Physics, Norwegian University of
	Science and Technology, N-7491 Trondheim, Norway}

\begin{abstract}
The electromagnetic response of topological insulators and superconductors is governed by a modified set of Maxwell equations that derive from a topological Chern-Simons (CS) term in the effective Lagrangian with coupling constant $\kappa$. Here we consider a topological superconductor or, equivalently, an Abelian Higgs model in $2+1$ dimensions with a global $O(2N)$ symmetry  
in the presence of a CS term, but without a Maxwell term.  
At large $\kappa$, the gauge field decouples from the complex scalar field, leading to a quantum critical behavior in the $O(2N)$ universality class. 
When the Higgs field is massive, the universality class is still governed by the $O(2N)$ fixed point. However, we show that the massless theory belongs 
to a completely different universality class, exhibiting an exotic critical behavior beyond the Landau-Ginzburg-Wilson paradigm.   
For finite $\kappa$ above a certain critical value $\kappa_c$, a quantum critical behavior with continuously varying critical exponents arises. However, as a function $\kappa$ a transition takes place for $|\kappa| < \kappa_c$ where conformality is lost.  Strongly modified scaling relations ensue. For instance, 
in the case where $\kappa^2>\kappa_c^2$, leading to the existence of a conformal fixed point, critical exponents are a function of 
$\kappa$.  
\end{abstract}


\maketitle

\section{Introduction}

\subsection{Conformal phase transition}

A conformal phase transition (CPT) \cite{miransky1997conformal} is defined as featuring a critical point with a non-power law diverging 
correlation length,  and which exhibits a universal jump in some generalized stiffness of the system. The cardinal  example of such a transition is 
the Berezinskii-Kosterlitz-Thouless (BKT) phase transition  \cite{berezinskii1971destruction,kosterlitz1973ordering} 
taking place in two-dimensional superfluids and superconductors when they transition from 
the low-temperature phase to the normal state, and in the melting transition of two-dimensional crystals \cite{kleinert1989gauge,nelson2002defects}.  
A key point of such phase transitions is the absence 
of a traditional Landau-type (local) order parameter with which to monitor the transition. 
In the above examples, the lack of a local order parameter is due to a fundamental theorem by 
Mermin and Wagner \cite{mermin1966absence}, which states that 
spontaneous breaking of continuous symmetries in two dimensions at any non-zero temperature cannot take place.  

The presence of 
a strongly fluctuating gauge-field puts an even stronger limitation on the existence of a local order parameter than the Mermin-Wagner theorem 
does. Namely,  in {\it any} gauge theory in {\it any dimension} such as for instance the Ginzburg-Landau theory of superconductors or, 
equivalently, the Abelian Higgs model (AHM) in 2+1 dimensions, an order parameter cannot be defined, unless it is gauge-invariant.
This result, known as Elitzur's theorem \cite{elitzur1975impossibility} implies that {\it no local order parameter exists for a superconductor}. 

On the other hand, response functions are gauge invariant. They are computed in terms of correlation functions of conserved currents. 
Response functions should exhibit universal features at a phase transition, provided the phase transition occurs at a critical point, 
that is, there exists a diverging length in the problem rendering the system scale-free at the transition. In an ordinary second-order 
phase transition,  with power-law divergence of some correlation length and susceptibilities, the universal aspects are associated 
with the exponents of the power laws. A much studied quantity both experimentally and theoretically is the current-current correlation 
function, which features an overall multiplicative constant, 
the superfluid density of the system. The universal aspect of this quantity at a 
standard second-order phase transition of a bulk superconductor is the exponent determining how the superfluid density vanishes as 
$T \to T_c$ from below. For thin film superconductors, the universal aspect of the same response function is a universal jump in the  
superfluid density at the transition, and a concomitant diverging correlation length with an essential singularity \cite{nelson1977universal}. 

\subsection{Conformality loss argument}

A general framework to derive BKT-like scaling in other theories was provided by Kaplan {\it et al.} \cite{kaplan2009conformality}
who showed that a CPT can be understood in terms of a conformality loss argument. A more mathematically precise discussion can be found 
in Ref. \cite{gorbenko2018walking}.  
Simply stated, it amounts to considering a renormalization group (RG) flow for a coupling $g$ depending on some 
parameter $\alpha$ such that $\beta(g;\alpha)\equiv \mu dg/d\mu=\alpha-\alpha_*-(g-g_*)^2$ \cite{kaplan2009conformality}. 
For $\alpha>\alpha_*$ the fixed points $g_\pm=g_*\pm\sqrt{\alpha-\alpha_*}$ are obtained, with $g_-$ infrared stable (IR) and 
$g_+$ ultraviolet (UV) stable.  Such a hypothetical RG flow describes a CPT as $\alpha$ is varied. Indeed, the IR and UV fixed points 
merge when $\alpha=\alpha_*$. But for $\alpha<\alpha_*$ conformality is lost, since $g_\pm$ become complex.  The BKT-like 
scaling follows easily by integrating the RG equation, yielding $\Lambda_{IR}/\Lambda_{UV}\approx \exp(-\pi/\sqrt{\alpha_*-\alpha})$ 
when $|g_{IR,UV}-g_*|\gg\sqrt{|\alpha-\alpha_*|}$. Recall that in the BKT scaling the inverse correlation length has the 
form, $\xi^{-1}\sim\exp(-{\rm const}/\sqrt{T-T_c})$ \cite{nelson1977universal}, so by comparison the parameter $\alpha$ plays a 
role analogous to the inverse temperature in the BKT transition 
\footnote{It is important to emphasize, however, that technically there are important distinctions between the BKT transition and 
	conformality loss, as discussed by Gorbenko {\it et al.} \cite{gorbenko2018walking}.}. 
This situation is also reminiscent of the one occuring in spinor QED in 2+1 dimensions (QED$_3$), 
where a gap generation occurs due to spontaneous chiral symmetry breaking, having 
the form, $m_{\rm QED}\sim\exp(-2\pi/\sqrt{N_c-N})$, where $N$ is the number of Dirac fermion species and 
$N_c=32/\pi^2$ \cite{appelquist1988critical}. This behavior of QED in 2+1 dimensions has been identified in Ref. \cite{gusynin1998effective} 
as a CPT. In this case the CPT is also a consequence of the non-locality of the Maxwell term at strong coupling, with the conformality lost 
point of view of merging of fixed points analyzed in Ref. \cite{Herbut-PhysRevD.94.025036}. 
A similar behavior leading to a CPT is also found in graphene in the presence of 
Coulomb interactions, where an excitonic gap is generated by spontaneous chiral symmetry breaking \cite{khveshchenko2001ghost}. 
Interestingly, in the case of graphene the CPT occurs as the coupling constant is varied, rather than the number of 
components \cite{gorbar2002magnetic}. This is due to the fact that in graphene the bare Coulomb interaction is three-dimensional 
(i.e., $\sim 1/r$) rather than logarithmic.  
Also in the context of so called deconfined quantum critical points 
\cite{senthil2004deconfined} a CPT may occur in an $N$-component AHM in 2+1 dimensions at low $N$ and in the strongly 
coupled regime \cite{nogueira2013deconfined}. 
In this case  when 
the number of components $N$ is varied below a certain critical value $N_c$ the fixed points become complex, resulting in conformality loss  \cite{nogueira2013deconfined,benvenuti2018qed}. Recently this behavior of the AHM has been addressed within the framework of the $\epsilon$-expansion 
up to four loops \cite{ihrig2019abelian}.  
In the past this behavior of the AHM was interpreted as a weak first-order phase 
transition \cite{HLM-PhysRevLett.32.292}. More recently the weak first-order phase transition in the Potts model with $Q>4$ has been also understood 
in terms of an approximate conformality loss due to its proximity to complex fixed points \cite{ConfLostPotts-PhysRevB.99.195130,Gorbenko-10.21468/SciPostPhys.5.5.050}. 

\subsection{Topological Abelian Higgs  model}

Here we introduce a more subtle type of CPT driven by a {\it topological} term in the effective action. The main 
motivation comes from the modification of Maxwell electrodynamics in topological materials 
\cite{Wen-Niu_PhysRevB.41.9377,Qi-2008,Qi-Witten-Zhang-PhysRevB.87.134519}.  For instance, the surface of a 
topological superconductor corresponds to an AHM in presence of a CS term \cite{Qi-Witten-Zhang-PhysRevB.87.134519}.  
In the absence of a Maxwell term, this model has a soliton solution in the form of a 
self-dual CS vortex \cite{frohlich1989quantum,Jackiw-Weinberg}. The Lagrangian  is simply given by, 
\begin{eqnarray}
\label{Eq:CSAHM}
\mathcal{L}&=&\frac{\kappa}{2}\epsilon_{\mu\nu\lambda}a^\mu\partial^\nu a^\lambda+|(\partial_\mu-ia_\mu)\phi|^2
\nonumber\\
&-&m_0^2|\phi|^2-\frac{u_0}{2}|\phi|^4,
\end{eqnarray} 
where $\kappa$ is the Chern-Simons coupling and $m_0^2$ and $u_0$ are written with a "0" subscript to emphasize that they represent bare quantities at this stage. Note that $a^\mu$ is a fluctuating field and not a background (external) gauge potential. 

In the limit $\kappa\to\infty$ the gauge field is frozen to zero and the theory becomes simply a globally $U(1)$-invariant scalar theory, which 
within an imaginary time formalism governs the 
universality class of a three-dimensional XY classical ferromagnet, which is the same as the universality class 
of superfluid Helium in three dimensions \cite{zinn1996quantum}. This theory is known to be exactly dual to an AHM without a CS term (note that in this 
case there is a Maxwell term) \cite{PESKIN1978122,THOMAS1978513,Dasgupta-Halperin_PhysRevLett.47.1556,kleinert1982disorder}.  
For $\kappa=1/(2\pi)$, corresponding to level 1 CS AHM,  
it has been recently pointed out in several papers that the Lagrangian (\ref{Eq:CSAHM}) maps via a bosonization duality to 
free Dirac fermions in 2+1 dimensions \cite{SEIBERG2016395,Karch_PhysRevX.6.031043,Mross_PhysRevX.7.041016,hsin2016level,komargodski2018symmetry,aharony2017chern,benini2018three,Raghu_PhysRevLett.120.016602,FERREIROS20181,NASTASE2018145}. The duality is assumed 
to be valid also when the fields are massless. Since a genuine duality is supposed to map a strongly coupled theory on one side to a weakly coupled 
theory on the other side, the statement we just made might at first sight sound confusing, since the theory on one side does not interact. Actually, it is 
the IR fixed point of the model (\ref{Eq:CSAHM}) that it is being mapped to the free fermion model.  In this paper we show that the IR behavior 
underlying the duality transformation is subtle in the massless regime of Eq. (\ref{Eq:CSAHM}), as the IR fixed point is dependent on the CS coupling 
$\kappa$ and that conformality of the IR fixed point is lost as $|\kappa|$ is varied below a certain critical value. However, this scaling regime 
typically corresponds to values of $\kappa$ larger than the one associated to the level 1 theory. On the other hand, we will show that the massive theory 
implies a scaling behavior featuring a Wilson-Fisher fixed point for all values of $\kappa$ and not only $\kappa=1/(2\pi)$. 
An immediate consequence of this result is that there must be two different paths to constructing the continuum limit of the theory (\ref{Eq:CSAHM}) 
from a lattice model. One example is provided by the bosonization duality derived using 
Wilson lattice fermions as discussed in Ref. \cite{Raghu_PhysRevLett.120.016602}. 

\section{Renormalization group for the massless theory}

\begin{figure}
	\setlength{\unitlength}{0.7mm}
	\includegraphics[width=2cm]{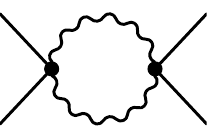}
	\caption{Photon bubble contributing to the effective Higgs interaction.}
	\label{Fig:fish}
\end{figure}

Fixing the Landau gauge, the one-loop Feynman diagram contributing to the effective Higgs $|\phi|^4$ coupling crucial for our analysis is 
given in Fig. \ref{Fig:fish}, which is proportional to $e^4I(p)$,
where $I(p)$ is a momentum space integral  in Euclidean spacetime (see Appendix \ref{App:CalcDiagrams}). The wiggles represent photon propagators, 
while the external legs are Higgs scalars.  
Here we assume the presence of an evanescent ($e^2\to\infty$) Maxwell term 
as a regulator term in order 
to allow for a better analysis of the interplay between IR and UV energy scales.  
In absence of a CS term $\lim_{|p|\to 0}I(p)$ is both IR and UV divergent, so usually we compute this 
diagram in this case assuming a nonzero momentum scale $|p|=\mu$ \cite{collins1985renormalization}. On the other hand, 
dimensional regularization \cite{Itzykson-Zuber}  
would in principle imply that  $\lim_{|p|\to 0}I(p)=0$, since such a regularization procedure usually 
compensates powers of IR and UV cutoffs \cite{collins1985renormalization}. Indeed, using cutoffs, we have $I(0)=2(\Lambda_{IR}^{-1}-\Lambda_{UV}^{-1})$ when 
$\kappa=0$, so $I(0)$ would vanish provided $\Lambda_{IR}=\Lambda_{UV}$. In the presence of the CS term, on the other hand, we 
have that $I(0)$ vanishes identically at fixed dimension $d=2+1$, being completely insensitive to IR and UV scales. 
Therefore,  we expect that $g(0)$ and $g(\mu)$ 
lead to very different fixed points. 

\begin{figure}
	\subfigure{\includegraphics[width=2cm]{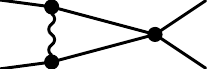}}
	\hfill
	\subfigure{\includegraphics[width=2cm]{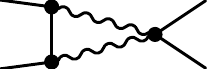}}
	\hfill
	\subfigure{\includegraphics[width=2cm]{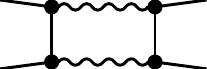}}
	\caption{Other diagrams containing photon lines which vanish in the Landau gauge.}
	\label{Fig:fish-1}
\end{figure}

For the case of a CS AHM (\ref{Eq:CSAHM}), the integral $I(p)$ is calculated explicitly in the Appendix  \ref{App:CalcDiagrams} 
assuming the Landau gauge, and 
is indeed found to vanish for $|p|\to 0$ for all $\kappa\neq 0$. More importantly, we have, 
\begin{equation}
\label{Eq:Vertex-photon-loop-infty}
\lim_{e^2\to\infty}e^4I(p)=-\frac{|p|}{8\kappa^2}.
\end{equation}
Thus, what in the usual (non-topological) AHM yields a term of order $e^4$ in a perturbation theory in terms of both $u$ and $e^2$, becomes here 
a term of order $1/\kappa^2$, reflecting a perturbation in powers of $1/|\kappa|$ instead. 
Note that the $e^2\to\infty$ limit yields a negative sign. This is to be contrasted to the $\kappa\to 0$ limit, 
\begin{equation}
\lim_{\kappa\to 0}e^4I(p)=\frac{3}{16|p|},
\end{equation}
producing a positive sign. This is an important point, as precisely the positive sign of this term prevents the existence of charged fixed points for 
a small number of Higgs fields in 2+1 dimensions \cite{HLM-PhysRevLett.32.292}.  

In order to put in perspective the role of the Maxwell term as a regulator in the AHM, let us assume the Lagrangian as it is given in Eq. (\ref{Eq:CSAHM}), i.e., 
without a Maxwell term, but with $e\neq 0$. In this case the propagator in the Landau gauge is given simply by, 
\begin{equation}
\label{Eq:CS-prop-Landau}
D_{\mu\nu}(p)=-\frac{1}{e^2\kappa}\frac{\epsilon_{\mu\nu\lambda}p_\lambda}{p^2},
\end{equation} 
leading to, 
\begin{equation}
\label{Eq:Fish-1}
e^4I(p)=\frac{2}{\kappa^2}\left[\int\frac{d^3q}{(2\pi)^3}\frac{1}{q^2}-p^2\int\frac{d^3q}{(2\pi)^3}\frac{1}{(p-q)^2q^2}\right].
\end{equation}
The first integral on the right-hand side of the  above equation is divergent and needs to be regularized, while the second integral is finite, yielding 
$1/(8|p|)$. We see that when a Maxwell term is present, $I(p)$ produces a finite result for all $e^2\neq 0$, including the limit case $e^2\to\infty$. 
On the other hand, with the propagator (\ref{Eq:CS-prop-Landau}) we can evaluate the divergent integral in Eq. (\ref{Eq:Fish-1}) with a cutoff and 
absorb the result into the bare coupling $u_0$. In this way we obtain the same result as before when the limit $e^2\to\infty$ is taken. 

The calculation using the propagator (\ref{Eq:CS-prop-Landau}) has the advantage of making calculations easier, even though sometimes an explicit 
UV cutoff is needed. For instance, it is more easily shown that the diagrams of Fig. \ref{Fig:fish-1} vanish in the Landau gauge even when a CS term 
is present. For the usual AHM it is a well known fact that  the diagrams of Fig. \ref{Fig:fish-1} 
vanish as $p\to 0$ in the Landau gauge \cite{Coleman-Weinberg_PhysRevD.7.1888}. 
in 2+1 dimensions the same is true using massless scalars at nonzero external momenta. Indeed, dimensional analysis implies that diagrams have a value 
$c/|p|$, where $c$ is some real constant. Since the diagrams must vanish in the $|p|\to 0$ limit, we must necessarily have $c=0$. The same behavior  
holds true for the model (\ref{Eq:CSAHM}). 

We may wonder about what happens if a different gauge, say, the Feynman gauge is chosen instead. In this case the calculations are longer, but gauge 
invariant results should not change. For example, while the wave function renormalization of the scalar field and the renormalization of the 
Higgs self-coupling are both gauge dependent, the RG $\beta$ functions are gauge independent (for an example, see Refs. \cite{Kang_PhysRevD.10.3455} 
and \cite{schakel1998boulevard}). 

Usually in order to obtain the RG $\beta$ function we need also the wave function 
renormalization $Z_\phi$ for the scalar field.  
Since $\kappa$ is scale invariant, $Z_\phi$ will at the end give no contribution to the one-loop $\beta$ function. 
The one-loop self-energy $\Sigma(p)$ excluding tadpole diagrams  satisfies, 
\begin{equation}
\lim_{e^2\to\infty}\Sigma(p)=-\frac{2p^2}{3\pi|\kappa|},
\end{equation}
and has this form even if $m\neq 0$ (see Appendix \ref{App:CalcDiagrams}).  
This implies,  
\begin{equation}
Z_\phi=\frac{1}{1-2/(3\pi|\kappa|)}.
\end{equation}
Note that due to the infrared bound \cite{froelich1976infrared}, $0<Z_\phi/p^2\leq 1/p^2$, we have necessarily 
the bound $|\kappa|>2/(3\pi)$. Thus, the limit $e^2\to\infty$ and quantum fluctuations at one-loop prevent us to allow $\kappa$ to vanish, 
although this is certainly possible at the classical level. 

\begin{figure}
	\includegraphics[width=4cm]{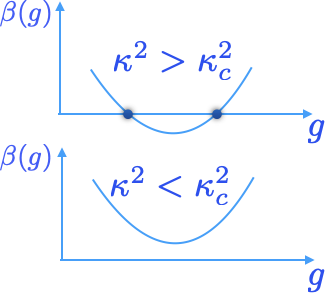}
	\caption{Schematic behavior of the RG $\beta$ function (\ref{Eq:beta-g-1}). Real fixed points exist only for $\kappa^2>\kappa_c^2$, leading to a 
		quantum critical behavior.  Conformality is lost for $|\kappa|<\kappa_c$ and a BKT-like quantum phase transition occurs in the topological superconductor.}
	\label{Fig:betafunc}
\end{figure}

The dimensionless effective coupling  is given by $g(\mu)=Z_\phi^2u(\mu)/\mu$, where $\mu=|p|$,  
and, 
\begin{equation}
u(\mu)=u_0-\left(\frac{N+4}{8}\right)\frac{u_0^2}{\mu}+\frac{\mu}{\kappa^2},
\end{equation}
corresponding to the sum of the one-loop diagrams with scalar and and photon bubbles.   
Thus, 
\begin{equation}
g(\mu)=\left(1+\frac{4}{3\pi|\kappa|}\right)\frac{u_0}{\mu}-\left(\frac{N+4}{8}\right)\frac{u_0^2}{\mu^2}+\frac{1}{\kappa^2},
\end{equation}
where we have assumed that there are $N$ complex scalar fields and that $1/\kappa^2$ and $u_0/\mu$ are of the same order. Therefore, the RG $\beta$ function,  
$\beta(g)=\mu dg/d\mu$
for the dimensionless coupling 
is given by (recall that $\kappa$ does not flow \cite{COLEMAN1985184}), 
\begin{eqnarray}
\beta(g)
&=&-\underbrace{\left[\left(1+\frac{4}{3\pi|\kappa|}\right)\frac{u_0}{\mu}-\left(\frac{N+4}{8}\right)\frac{u_0^2}{\mu^2}+\frac{1}{\kappa^2}\right]}_{=g}
\nonumber\\
&+&\left(\frac{N+4}{8}\right)\frac{u_0^2}{\mu^2}+\frac{1}{\kappa^2}.
\end{eqnarray}
The above equation can be rewritten within the accuracy of the one-loop approximation as, 
\begin{equation}
\label{Eq:beta-g-1}
\beta(g)=\frac{g_*}{2}\left[\frac{\kappa_c^2}{\kappa^2}-1+\left(\frac{g}{g_*}-1\right)^2\right],
\end{equation}
where $g_*=4/(N+4)$ and $\kappa_c^2=2/g_*$.  The RG $\beta$ function (\ref{Eq:beta-g-1}) has precisely the paradigmatic form 
discussed by Kaplan {\it et al.} \cite{kaplan2009conformality} for theories featuring conformality lost. The only difference is that in our case 
the $\beta$ function has the opposite sign. Depending on the range of $\kappa$, the theory may have a conformal fixed point or not. The $\beta$ 
function profile is shown schematically in Fig. \ref{Fig:betafunc}. 

Nontrivial fixed points corresponding to quantum criticality exist whenever 
$\kappa^2\geq\kappa_c^2$. 
We note that in contrast to the usual AHM in 2+1 dimensions \cite{HLM-PhysRevLett.32.292}, the CS AHM features a quantum critical point for all $N$ if the CS coupling satisfies the inequality $\kappa^2\geq\kappa_c^2$.  Indeed, this regime features the IR ($g_+$) and UV ($g_-$) stable fixed points, 
\begin{equation}
\label{Eq:FPs}
g_\pm=g_*(1\pm\sqrt{1-\kappa_c^2/\kappa^2}),
\end{equation}
 with  the IR stable fixed point corresponding to the quantum critical point of the theory. 
 
 In order for perturbation theory to be well controlled we need a parameter to guarantee the smallness of both $g$ and $1/\kappa^2$. As usual, 
 such a smallness is dictated by the fixed point structure of the theory. Since $g_\pm\propto g_*$, we have that the fixed points are small for a large enough 
 value of $N$. Indeed, for $N$ large $g_*\sim\mathcal{O}(1/N)$. Thus, for $g $ not too far from $g_\pm$, perturbation theory is well behaved. 
 Similarly, we have that $1/\kappa^2\sim\mathcal{O}(1/N)$ for $\kappa$ near $\kappa_c$.  We can in principle also 
 speculate that even for $N=1$ perturbation theory is well controlled, since we still have $g_*<1$. However, without a careful large order behavior 
 analysis, such a claim remains inconclusive.

Conformality is lost when $|\kappa|<\kappa_c$, corresponding to the situation where the above fixed points become complex. 
The solution of the differential equation (\ref{Eq:beta-g-1}) for this case is, 
 \begin{eqnarray}
 \label{Eq:Sol-beta}
 \ln\left(\frac{\mu}{\Lambda}\right)&=&\frac{-2}{\sqrt{\kappa_c^2/\kappa^2-1}}\left[\arctan\left(\frac{1-g/g_*}{\sqrt{\kappa_c^2/\kappa^2-1}}\right)
 \right.\nonumber\\
 &+&\left.\arctan\left(\frac{g_\Lambda/g_*-1}{\sqrt{\kappa_c^2/\kappa^2-1}}\right)\right],
 \end{eqnarray}
 where $g_*<g_\Lambda=g(\Lambda)$.  When $\kappa\to\kappa_c$ for $|\kappa|<\kappa_c$ 
 the complex fixed points merge and we obtain that for $|g_\Lambda/g_*-1|\gg\sqrt{\kappa_c^2/\kappa^2-1}$ the momentum 
 scale satisfies, 
  \begin{equation}
 \label{Eq:ConfLost}
 \frac{\mu}{\Lambda}=
 \exp\left(-\frac{\pi+\theta(\kappa,g)}{\sqrt{\kappa_c^2/\kappa^2-1}}\right), 
 \end{equation}
where $\theta(\kappa,g)=2\arctan[(1-g/g_*)/\sqrt{\kappa_c^2/\kappa^2-1}]$, implying a BKT-like scaling when $g<g_*$. 
On the other hand, as $\kappa\to\kappa_c$, we obtain for all $g>g_*$, 
\begin{equation}
\label{Eq:Sing-g}
\frac{\mu}{\Lambda}\mathop{=}_{~~\kappa\to\kappa_c}\exp\left(-\frac{2g_*}{g-g_*}\right),
\end{equation}
which features an essential singularity at $g=g_*$, representing 
a behavior similar to the one obtained in the case of deconfined quantum critical points \cite{nogueira2013deconfined}. 
Therefore, when $\kappa\in[-\kappa_c,\kappa_c]$ 
Eqs. (\ref{Eq:ConfLost}) and (\ref{Eq:Sing-g}) imply that only for $g< g_*$ the system becomes critical as $\kappa$ approaches 
$\kappa_c$, implying a BKT-like critical point. For $g>g_*$ the system does not become critical as $\kappa\to\kappa_c$, needing in addition that 
$g\to g_*$, corresponding to the fixed point in this case. 

The solution (\ref{Eq:Sol-beta}) can also be used for $\kappa^2>\kappa_c^2$ after performing a simple manipulation with complex numbers. In 
this case we obtain a quantum critical scaling behavior,  
\begin{equation}
\label{Eq:QC-scaling}
\frac{\mu}{\Lambda}\approx\left(\frac{g_+-g}{g-g_-}\right)^{1/\sqrt{1-\kappa_c^2/\kappa^2}}, 
\end{equation}
where we have assumed once more that $|g_\Lambda/g_*-1|\gg\sqrt{1-\kappa_c^2/\kappa^2}$. The above solution makes it apparent that  
$g=g_+$ is an IR stable fixed point corresponding to $\mu/\Lambda\to 0$, while $g_-$ is a UV stable fixed point corresponding to $\mu/\Lambda\to\infty$. 
As $\kappa\to\kappa_c$ leading to a merging of the IR and UV stable fixed points,  
Eq. (\ref{Eq:QC-scaling}) becomes Eq. (\ref{Eq:Sing-g}) and the system undergoes a CPT. 

\section{Renormalization group for the massive theory}

For a nonzero renormalized mass $m$ the one-loop scalar field bubble diagram yielding the $u_0^2$ contribution for the vertex function does not diverge 
for $\mu=|p|\to 0$. Thus, we can use $m$ as RG scale instead \cite{zinn1996quantum,parisi1980field}. However, as we have already 
seen, in this case the photon bubble in Fig. \ref{Fig:fish} vanishes identically. Furthermore, due to the $e^2\to\infty$ limit the wavefunction 
renormalization of the Higgs field does not contribute to $\beta(g)=mdg/dm$, just as before, since $\kappa$ itself does not flow. As a result, the critical 
behavior as $m\to 0$ is governed by the Wilson-Fisher fixed point. We conclude therefore that in the critical behavior of the topological 
AHM the limits $\mu\to 0$ and $m\to 0$ do not commute. Thus, the critical behavior of the massive theory does not feature conformality lost as $\kappa$ is 
varied. This lack of commutativity in the scaling behavior is a unique feature of interacting CS field theories. Such a behavior is  
more explicit in the CS term itself, when the latter is generated by quantum fluctuations after integrating out Dirac fields in 2+1 
dimensions \cite{CS,Redlich,Semenoff_PhysRevLett.63.2633,Nogueira-Eremin-PhysRevB.88.085126}. In that case the odd parity contribution 
to the vacuum polarization yielding the CS term is only nonzero if the Dirac fermion is massive. However, after performing the loop integral the 
end result depends only on the sign of the mass $M$ of the Dirac field, corresponding to a CS coupling  
$\kappa=(4\pi)^{-1}M/|M|$ \cite{Redlich,Semenoff_PhysRevLett.63.2633,Nogueira-Eremin-PhysRevB.88.085126}. Thus, the CS term survives 
the $M\to 0$ limit {\it after} the quantum fluctuations are calculated. 
Unlike the fermionic case, the mass $m$ of the Higgs field has nothing to do with the presence or 
absence of a CS term in the Lagrangian. Nevertheless, a nonzero $m$ has an indirect relation to the scaling behavior in an AHM with a CS term, 
due to the vanishing of the diagram Fig. \ref{Fig:fish} as $|p|\to 0$.

\section{Superfluid stiffness} 
On the basis of the above we now provide a concrete and in principle testable prediction on the behavior of the superfluid stiffness in the topological Higgs superconductor.
The superfluid stiffness $\rho_s$ is a response function given quite generally 
by the current correlation function at zero momentum \cite{Weichman_PhysRevB.38.8739}. 
Its scaling behavior is given 
 Josephson scaling relation, $\rho_s\sim(T_c-T)^{\nu(D-2)}$ \cite{JOSEPHSON1966608}, 
where $D$ is the dimension of space.  In the two-dimensional case, the result of Ref. \onlinecite{JOSEPHSON1966608} immediately implies a jump as $T_c$ is approached from the left, 
since the superfluid stiffness must vanish for $T>T_c$. A hallmark of the transition is that this jump is universal  \cite{nelson1977universal}.   

In the present case it is not the dimension of space that is relevant in the scaling of the stiffness, but rather the dimension of spacetime, $d=2+1$, which in 
the context of quantum critical phenomena can be regarded as a theory with dynamical exponent $z=1$ \cite{Sachdev-book}, 
corresponding to a Lorentz-invariant system. The role of the temperature is played by the bare mass squared, $m_0^2$, which in the massless case is tuned to a critical value 
$m_{0c}^2$. In the massless case 
we obtain that the critical exponent $\nu$ is defined as usual via insertions of the operator $|\phi|^2$, whose anomalous dimension is 
 $3-1/\nu$ \cite{zinn1996quantum}.  
Clearly, this exponent is only defined for $\kappa^2\geq \kappa_c^2$, with the result, 
\begin{equation}
\label{Eq:nu}
\nu(\kappa,N)=\frac{2}{4-\left(\frac{N+1}{N+4}\right)(1+\sqrt{1-\kappa_c^2/\kappa^2})}.
\end{equation}
We see that in this case the critical exponent is not a number, but a function of the CS coupling $\kappa$. Note that for $\kappa\to\infty$ it agrees with 
the one-loop result for a $O(2N)$ classical Heisenberg model, as expected, since for $\kappa\to\infty$ the gauge field and the scalar field 
decouple. For the $U(1)$ symmetric case ($N=1$) we obtain $\nu=5/9$ at the critical value $\kappa=\kappa_c$. 

Continuously varying critical exponents is a well known feature of some CS theories.     
A closely related model where this occurs is the CP$^{N-1}$ model 
with a CS term, which has been studied in detail for large $N$ \cite{Ferretti-Rajeev}.   Note that in the model 
we have considered  the massless regime does not smoothly connect to larger values of $N$, since in the large $N$ limit  
$\kappa_c^2$ becomes large and conformality is lost. The large $N$ results of Ref. \cite{Ferretti-Rajeev} were 
obtained in the massive regime implied by the  CP$^{N-1}$ constraint. We have seen that in the massive regime 
the model flows to a conformal fixed point. 

When a CPT occurs  the critical behavior of the stiffness is highly unusual due to the BKT scaling (\ref{Eq:ConfLost}). 
Because the theory is $2+1$-dimensional, the argument of the BKT universal jump in the superfluid stiffness is not exactly the same as in the case of a 
BKT transition \cite{nelson1977universal}. In fact,  the stiffness must scale as in Eq. (\ref{Eq:ConfLost}), meaning that it 
vanishes continuously as $\kappa\to\kappa_c$ if $g<g_*$. However, a jump would occur for $g>g_*$,  since for $\kappa\to\kappa_c$  
Eq. (\ref{Eq:Sing-g}) holds.  

\section{Discussion}
\label{Sect:Discussion}

Before concluding, it is worth putting into proper context the results we have found so far, especially due to the interest of the model within the framework 
of the so called "duality web" \cite{SEIBERG2016395}. While the RG result for massive scalars seems to be largely consistent with the bosonization 
duality scenario, the RG result for the massless case does not give conclusive answers, as the case $N=1$ is not fully controlled perturbattively. 
In fact, $\kappa=1/(2\pi)$ is smaller than $\kappa_c=\sqrt{(N+4)/2}$ for $N=1$. Thus, even if we assume the validity of perturbation theory down 
to the case of a single scalar field, the level 1 CS theory would be in the regime of loss of conformality. On the other hand, the RG analysis raises 
interesting questions for massless scalars in the large $N$ limit. First, note that the critical behavior clearly does not correspond to a Wilson-Fisher 
fixed point, although it is close to it for large $N$. This is in stark contrast to the massive case, where the Wilson-Fisher fixed point governs the 
critical behavior with $\kappa$ having an arbitrary value.

There are several details about the duality scenario that further complicates  the analysis. Although this is beyond the scope of the present paper, let us mention some of the issues. First, it must be noted that $N>1$ scalars necessarily implies a number of fermionic fields $N_f\neq 1$ and eventually a non-Abelian structure for the CS terms on each side of the duality \cite{hsin2016level,komargodski2018symmetry,aharony2017chern}.  In this case, the phase structure 
may exhibit a confining phase depending on the value of $N_f$. The second point, related to the first, is that we are attempting   to 
draw a comparison to the simpler $N_f=N=2\pi\kappa=1$ case \cite{SEIBERG2016395}, based on an RG analysis of 
a theory having a global $O(2N)$ symmetry and a local $U(1)$ one.  Here, we must note that a standard boson-boson 
duality of the type employed in lattice gauge theories is not known for the $O(2N)$ group. However, we are seeking a duality mapping fixed points theories, especially a situation where one of the sides of the duality is a free fermion theory. The best scenario would be to find a tractable interacting fermion theory whose fixed point and current correlations match the corresponding ones of the boson theory.

\section{Conclusions} 
We have shown that AHM with a CS term exhibits a much more peculiar quantum critical scaling behavior than has been realized previously. We have seen that the massless theory exhibits  quantum critical behavior with power law scaling of physical quantities only for a CS coupling $\kappa$ above a certain critical value $\kappa_c$. Although the critical behavior is governed by an IR stable fixed point leading to power law behavior, quantum criticality is highly unconventional, since critical exponents are a function of $\kappa$.   
For a CS coupling  below the critical value $\kappa_c$ there is a phase transition to a state featuring complex fixed points and 
conformality lost. The RG scale exhibits a BKT-like scaling in this case. On the other hand, if the model is massive and the critical point is 
approached by sending the mass to zero, conventional critical behavior with a Wilson-Fisher fixed point is obtained. Thus, the two limits of vanishing mass and momenta do not commute, leading to radically distinct forms of quantum criticality. 

The results we have obtained are relevant in light of recently well studied bosonization duality in $2+1$ dimensions 
\cite{SEIBERG2016395,Karch_PhysRevX.6.031043,Mross_PhysRevX.7.041016,hsin2016level,komargodski2018symmetry,aharony2017chern,benini2018three,Raghu_PhysRevLett.120.016602,FERREIROS20181,NASTASE2018145}. 
While this 
bosonization duality seems to be well established in the massive case, it remains a conjecture in the massless case. The unconventional 
criticality of the massless case shows that in the conformality lost regime, a duality to free massless Dirac fermions is unlikely, as the bosonic 
theory features complex fixed points.  However, as mentioned in Sect. \ref{Sect:Discussion}, our RG analysis of the massless case is essentially 
valid at large $N$, in a regime where the duality is anyway more complex \cite{komargodski2018symmetry}.

\acknowledgments
This work is supported by the DFG through the W\"urzburg-Dresden Cluster of Excellence on Complexity and Topology in Quantum Matter -- \textit{ct.qmat} (EXC 2147, project-id 39085490) and through SFB 1143 (project-id 247310070). Support from the Norwegian Research Council through Grant No
262633 ``Center of Excellence on Quantum Spintronics'', and Grant No. 250985, ``Fundamentals of Low-dissipative Topological Matter'' is acknowledged.

\appendix
\section{Calculation of integrals}
\label{App:CalcDiagrams}

\subsection{Calculation of $I(p)$}

For the sake of convenience, here we will set $M=e^2|\kappa|$. 
The Feynman diagram from Fig. \ref{Fig:fish} is proportional to the integral, 
\begin{equation}
\label{Eq:I}
I(p)=\int\frac{d^3k}{(2\pi)^3}D_{\mu\nu}(p-k)D_{\mu\nu}(k),
\end{equation} 
where,
\begin{equation}
D_{\mu\nu}(p)=\frac{1}{p^2+M^2}\left(\delta_{\mu\nu}-\frac{p_\mu p_\nu}{p^2}-\frac{M}{p^2}\epsilon_{\mu\nu\lambda}
p_\lambda\right),
\end{equation}
is the propagator in the Landau gauge. 

After performing the straightforward indices contraction in the integral (\ref{Eq:I}), we obtain, 
\begin{eqnarray}
I(p)&=&\int\frac{d^3q}{(2\pi)^3}\frac{1}{[(p-q)^2+M^2](q^2+M^2)}\nonumber\\
&\times&\left\{1+\frac{[q\cdot(p-q)]^2}{(p-q)^2q^2}+\frac{2M^2q\cdot(p-q)}{q^2(p-q)^2}
\right\}.
\end{eqnarray}

In what follows we use a series of simple algebraic manipulations to reduce $I$ to a combination of integrals, 
\begin{equation}
I_0=\int\frac{d^3q}{(2\pi)^3}\frac{1}{(p-q)^2q^2},
\end{equation}
\begin{eqnarray}
I_1&=&\int\frac{d^3q}{(2\pi)^3}\frac{1}{(p-q)^2(q^2+M^2)}
\nonumber\\
&=&\int\frac{d^3q}{(2\pi)^3}\frac{1}{[(p-q)^2+M^2]q^2},
\end{eqnarray}
\begin{equation}
I_2=\int\frac{d^3q}{(2\pi)^3}\frac{1}{[(p-q)^2+M^2](q^2+M^2)},
\end{equation}
which can be solved using the method of Feynman parameters \cite{Itzykson-Zuber} in a standard way to give, 
\begin{equation}
I_0=\frac{1}{8|p|},
\end{equation}
\begin{equation}
I_1=\frac{1}{4\pi|p|}\arctan\left(\frac{|p|}{|M|}\right),
\end{equation}
\begin{equation}
I_2=\frac{1}{4\pi|p|}\arctan\left(\frac{|p|}{2|M|}\right).
\end{equation}
The reduction of $I(p)$ to a combination of the above integrals is achieved by means of simple algebraic ticks, for instance, by using repeatedly relations like, 
\begin{eqnarray}
&&q\cdot(p-q)=\frac{p^2-q^2-(p-q)^2}{2}
\nonumber\\
&=&M^2+\frac{p^2-(q^2+M^2)-[(p-q)^2+M^2]}{2},
\end{eqnarray}
and, 
\begin{equation}
\frac{1}{q^2(q^2+M^2)}=\frac{1}{M^2}\left(\frac{1}{q^2}-\frac{1}{q^2+M^2}\right),
\end{equation}
in which case we obtain, 
\begin{eqnarray}
\label{Eq:Vertex-photon-loop}
e^4I(p)&=&e^4\left[\frac{1}{e^2|\kappa|}+\frac{|p|}{8e^4\kappa^2}\left(\frac{p^2}{4e^4\kappa^2}-1\right)\right.\nonumber\\
&+&\frac{1}{4\pi|p|}\left(-\frac{5}{2}+\frac{p^2}{e^4\kappa^2}-\frac{p^4}{2e^8\kappa^4}\right)\arctan\left(\frac{|p|}{e^2|\kappa|}\right)
\nonumber\\
&+&\left.\frac{p^4+16e^8\kappa^4}{16\pi e^8\kappa^4|p|}\arctan\left(\frac{|p|}{2e^2|\kappa|}\right)
\right],
\end{eqnarray}
It is easily obtained that, 
\begin{equation}
\lim_{|p|\to 0}I(p)=0, 
\end{equation}
for all $\kappa\neq 0$. On the other hand, we have, 
\begin{equation}
\lim_{\kappa\to 0}I(p)=\frac{3}{8|p|}. 
\end{equation}
Furthermore, we have, 
\begin{equation}
\lim_{e^2\to\infty}e^4I(p)=-\frac{|p|}{8\kappa^2},
\end{equation}
corresponding to Eq. (\ref{Eq:Vertex-photon-loop-infty}). 

\subsection{Calculation of $\Sigma(p)$}

The self-energy  $\Sigma(p)$, excluding tadpole diagrams, is given by, 
\begin{eqnarray}
\label{Eq:Sigma-int}
\Sigma(p)&=&-e^2\int\frac{d^3q}{(2\pi)^3}\frac{(2p_\mu-q_\mu)(2p_\nu-q_\nu)}{[(p-q)^2+m^2](q^2+e^4\kappa^2)}
\nonumber\\
&\times&\left(\delta_{\mu\nu}-\frac{q_\mu q_\nu}{q^2}
-\frac{e^2\kappa\epsilon_{\mu\nu\lambda}q_\lambda}{q^2}
\right)
\nonumber\\
&=&4e^2\int\frac{d^3q}{(2\pi)^3}\frac{(q\cdot p)^2/q^2-p^2}{[(p-q)^2+m^2](q^2+e^4\kappa^2)}.
\end{eqnarray} 
and involves two integrals, namely, 
\begin{equation}
J_1=\int\frac{d^3q}{(2\pi)^3}\frac{1}{[(p-q)^2+m^2](q^2+M^2)},
\end{equation}
and, 
\begin{equation}
J_2=\int\frac{d^3q}{(2\pi)^3}\frac{(2p\cdot q)^2}{[(p-q)^2+m^2](q^2+M^2)q^2}.
\end{equation}
The first integral is easily calculated with the method of Feynman parameters \cite{Itzykson-Zuber}, yielding, 
\begin{eqnarray}
J_1&=&\frac{1}{8\pi|p|}\left[\arctan\left(\frac{p^2+m^2-M^2}{2|M||p|}\right)
\right.\nonumber\\
&+&\left. \arctan\left(\frac{p^2+M^2-m^2}{2|m||p|}\right)
\right].
\end{eqnarray}
The calculation of $J_2$ takes more time, but it is also straightforward. First we rewrite it as, 
\begin{eqnarray}
J_2&=&\int\frac{d^3q}{(2\pi)^3}\frac{(2p\cdot q)[p^2+q^2-(p-q)^2]}{[(p-q)^2+m^2](q^2+M^2)q^2}
\nonumber\\
&=&\frac{2(p^2+m^2)}{M^2}\int\frac{d^3q}{(2\pi)^3}\frac{q\cdot p}{[(p-q)^2+m^2 ]q^2}
\nonumber\\
&-&\frac{2(p^2+m^2-M^2)}{M^2}\int\frac{d^3q}{(2\pi)^3}\frac{q\cdot p}{[(p-q)^2+m^2 ](q^2+M^2)}.
\nonumber\\
\end{eqnarray}
Now the fastest way to proceed is to use the method of Feynman parameters once more to obtain, 
\begin{eqnarray}
&&\int\frac{d^3q}{(2\pi)^3}\frac{q_\mu}{[(p-q)^2+m^2 ](q^2+M^2)}
\nonumber\\
&=&\frac{p_\mu}{8\pi}\int_{0}^{1}\frac{d\alpha~\alpha}{
	\alpha(1-\alpha)p^2+\alpha m^2+(1-\alpha)M^2}
\nonumber\\
&=&\frac{p_\mu}{16\pi|p|^3}\left\{2|p|(|M|-|m|)+(p^2+m^2-M^2)
\right.\nonumber\\
&\times&\left[
\arctan\left(\frac{p^2+m^2-M^2}{2|M||p|}\right)
\right.\nonumber\\
&+&\left.\left.\arctan\left(\frac{p^2+M^2-m^2}{2|m||p|}\right)
\right]
\right\}.
\nonumber\\
\end{eqnarray}
An integral corresponding to the limit $|M|\to 0$ of the above result is also needed in the expression for $J_2$. After carrying out some 
straightforward simplifications, we obtain, 
\begin{eqnarray}
\label{Eq:Sigma-Higgs}
\Sigma(p)&=&\frac{e^2}{4\pi}\left\{e^2|\kappa|-|m|-\frac{p^2+m^2}{e^2|\kappa|}
\right. \nonumber\\
&+&\frac{(p^2+m^2)^2}{2e^4\kappa^2|p|}\left[\frac{\pi}{2}+\arctan\left(\frac{p^2-m^2}{2|m||p|}\right)\right]
\nonumber\\
&-&\frac{(p^2+m^2-e^4\kappa^2)^2+4e^4\kappa^2p^2}{2e^4\kappa^2|p|}
\nonumber\\
&\times&\left[\arctan\left(\frac{p^2+e^4\kappa^2-m^2}{2|m||p|}\right)
\right.\nonumber\\
&+&\left.\left.\arctan\left(\frac{p^2+m^2-e^4\kappa^2}{2e^2|\kappa||p|}\right)
\right]
\right\}.
\end{eqnarray}
The wavefunction renormalzation is obtained by expanding $\Sigma(p)$ up to $p^2$, 
\begin{equation}
\Sigma(p)=-\frac{2e^2}{3\pi}\frac{p^2}{m+e^2|\kappa|}+{\cal O}(p^4),
\end{equation}
and we see that, 
\begin{equation}
\lim_{m\to 0}\left.\frac{\partial\Sigma}{\partial p^2}\right|_{p=0}=\lim_{e^2\to \infty}\left.\frac{\partial\Sigma}{\partial p^2}\right|_{p=0}=
-\frac{2}{3\pi|\kappa|},
\end{equation}
as asserted in the main text.  Furthermore, we note that, 
\begin{equation}
\lim_{e^2\to \infty}\Sigma(p)=-\frac{2p^2}{3\pi|\kappa|}.
\end{equation}

\section{Renormalized mass}

For completeness we give here the expression for the renormalized mass, which in the main text is assumed to vanish, 
\begin{eqnarray}
\label{Eq:Renorm-m2}
m^2&=&m_0^2+(N+1)u_0\int\frac{d^3q}{(2\pi)^3}\frac{1}{q^2+m^2}
\nonumber\\
&+&2e^2\int\frac{d^3q}{(2\pi)^3}\frac{1}{q^2+e^4\kappa^2}
\nonumber\\
&=&m_0^2+\frac{(N+1)u_0+2e^2}{2\pi^2}\Lambda
\nonumber\\
&-&\frac{(N+1)u_0|m|+2e^4|\kappa|}{4\pi},
\end{eqnarray}
where $\Lambda=\Lambda_{UV}$ and $N$ is the number of complex scalars. 
Thus, we see that the bare mass $m_0$ has to be chosen in such a way as to have a finite renormalized mass 
as $e^2\to\infty$. We might be worry that this is a somewhat artificial fine-tuning. However, we should note that $e^2$ actually behaves as a 
UV cutoff scale and can be considered as such. 

\bibliography{tcpt}

\end{document}